# An Introduction to Artificial Intelligence and Solutions to the Problems of Algorithmic Discrimination

*Nicholas Schmidt and Bryce Stephens*

*Nicholas Schmidt (nschmidt@bldsllc.com) is the AI Practice Leader at BLDS, LLC. He heads the artificial intelligence and machine learning innovation practice and specializes in the application of statistics and economics to questions of law, regulatory compliance, and best practices in model governance.*

*Bryce Stephens (bstephens@bldsllc.com) is a Director at BLDS, LLC. He provides economic research, econometric analysis, and compliance advisory services, with a specific focus on issues related to consumer financial protection, such as the Equal Credit Opportunity Act (ECOA), and emerging analytical methods.*

## I. Introduction

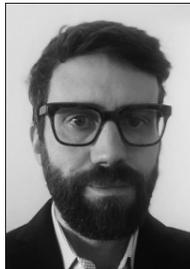
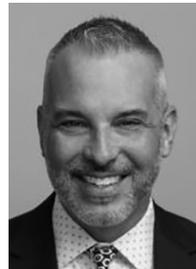

**Nicholas Schmidt**   **Bryce Stephens**

As artificial intelligence (AI) and machine learning (ML) have become more common in our daily lives, one of the most pressing issues in its adoption has been the difficulty in ensuring fairness and transparency in their use. There is substantial evidence that AI and ML algorithms can cause bias against minorities, women, and other protected classes. Facial recognition programs are more likely to misidentify people with darker skin; an employment test sometimes gives lower scores to women who graduated from women's colleges or who participated in women's sports; people of color may be less likely to receive certain marketing offers (including housing) because of social media behavior or geography; and minority groups have been shown to be falsely flagged as being more likely to be recidivists than whites.[1]

---

1. Jeffrey Dastin, *Amazon Scraps Secret AI Recruiting Tool that Showed Bias Against Women*, Reuters (Oct. 8, 2018), https://reut.rs/2Po4ZJi [https://perma.cc/K9GT-HBHR]; Julie Angwin & Jeff Larson, *Bias in Criminal Risk Scores is Mathematically Inevitable, Researchers Say*, ProPublica (Dec. 30, 2016), https://bit.ly/2iTc4B9 [https://perma.cc/PEL9-ZTHP]; Jeff Larson et al., *How We Analyzed the COMPAS Recidivism Algorithm*, ProPublica (May 23, 2016), https://bit.ly/1TGK42v [https://perma.cc/D3QJ-4M84]; Steve Lohr, *Facial Recognition is Ac-*

The increasing use of AI and ML techniques in employment, consumer finance, and housing has caught the attention of the federal regulatory community, the executive branch, and Congress. There is growing interest in these technological innovations—as well as the use of so-called alternative data—and the risks and benefits they pose. The White House, Consumer Financial Protection Bureau (CFPB), Federal Reserve Board, Federal Trade Commission, and the Government Accountability Office have all released reports or posted requests for information on topics related to algorithmic fairness; the use of AI and ML in employment, credit, and housing markets; and the use of alternative data.[2] Recently, Senators Booker, Wyden, and Clark introduced the Algorithmic Accountability Act, "which requires companies to study and fix flawed computer algorithms that result in inaccurate, unfair, biased or discriminatory decisions impacting Americans."[3]

Government interest has been focused on the potential benefits and risks of the use of AI, ML, and alternative data to consumers, employees, and markets. In credit markets, for example, there is hope that reliance on nontraditional credit data may lead to an expansion in access to credit for consumers who are currently not in the traditional credit system. Moreover, ML and AI techniques can be applied to large-scale datasets (e.g., big data)[4] and are better at predicting outcomes than traditional statistical techniques. Better model predictions lead to more efficient allocations in credit, labor, and housing markets.

The risks associated with AI, ML, and alternative data are most apparent when considering the contexts in which models and data are developed and used. While the use of AI opens new avenues of risk for consumers where additional legislation may be required (especially relating to pri-

---

*curate, if You're a White Guy*, N.Y. Times (Feb. 9, 2018), https://nyti.ms/2H3QeaT [https://perma.cc/J86U-89MQ].

2. U.S. Gov't Accountability Office, GAO–19–111, Financial Technology: Agencies Should Provide Clarification on Lenders' Use of Alternative Data (2018), https://bit.ly/2NBFBjm [https://perma.cc/VHE2-ML8Q]; Request for Information Regarding Use of Alternative Data and Modeling Techniques in the Credit Process, 82 Fed. Reg. 11183 (Feb. 21, 2017); Exec. Office of the President, Big Data: A Report on Algorithmic Systems, Opportunity, and Civil Rights (2016), https://bit.ly/2lCSs8H [https://perma.cc/ZL8V-2PKF]; Fed. Trade Comm'n, Big Data: A Tool for Inclusion or Exclusion? Understanding the Issues (2016), https://bit.ly/1n52gG6 [https://perma.cc/MW9H-VWGQ]; Carol A. Evans, *Keeping Fintech Fair: Thinking About Fair Lending and UDAP Risks*, Consumer Compliance Outlook (2017), https://bit.ly/2HaqfR1 [https://perma.cc/8UJN-X7GY].

3. Press Release, Cory Booker for U.S. Senator for N.J., Booker, Wyden, Clarke Introduce Bill Requiring Companies to Target Bias in Corporate Algorithms (Apr. 10, 2019), https://www.booker.senate.gov/?p=press_release&id=903 [https://perma.cc/N29T-9SQM].

4. *See* discussion *infra* Section III.

vacy), there are already a number of federal and many state-level laws addressing discrimination and unfairness that can be applied to the outcomes driven by these technologies. Specifically, federal laws have been enacted to protect consumers from discrimination in credit, housing, and employment, where federal regulators and agencies are tasked with enforcing these laws.[5] Additionally, there are laws in place to ensure that consumers understand why they are denied access to credit and to ensure the right for consumers to correct errors in their credit reports.[6] There are also federal and state laws that protect consumers from unfair and deceptive practices.[7]

Despite the laws and regulations that are already in place, the advent of AI, ML, and alternative data in automated decision-making poses unique challenges for regulators, lenders, and firms when evaluating models for compliance and potential violations of consumer protection laws. Models that are built using ML or more advanced AI techniques often exhibit a black box quality, meaning that the relationships between model inputs and model predictions may not be straightforward or easy to describe. This makes explaining model outcomes a challenging, though no longer intractable, problem. Developing approaches to minimize discriminatory impact is also more challenging with ML and AI methods. Finally, reliance on alternative data raises many questions around considerations of compliance and policy. For instance, if a consumer's educational attainment is predictive of credit default, but also highly correlated with race, can it be used in a credit model without running afoul of the Equal Credit Opportunity Act's (ECOA) anti-discrimination provisions? If so, would relying on educational attainment data perpetuate, or worse, amplify historical inequities in access to credit?[8]

---

5. Federal anti-discrimination laws include: the Equal Credit Opportunity Act (ECOA) (consumer credit), the Fair Housing Act, and Title VII (employment). 15 U.S.C. § 1691 (2017); 42 U.S.C. § 3604; *id.* § 2000(e).

6. The Fair Credit Reporting Act (FCRA) governs how consumer credit information is collected, maintained, and used. 15 U.S.C. § 1681. Both the FCRA and the ECOA require that a consumer be made aware of the principal reasons for a denied request for credit. *Id.* §§ 1681, 1691.

7. There are two federal statutes: Unfair or Deceptive Acts or Practices and Unfair, Deceptive, or Abusive Acts and Practices. *Id.* § 45; 12 U.S.C. § 5531. Most states and United States territories have adopted statutes mirroring these federal statues.

8. The example raised is not hypothetical. In 2017, Upstart, Inc. applied for and received a no action letter from the Consumer Financial Protection Bureau indicating that the Bureau would take no supervisory or enforcement action with respect to potential violations of the ECOA for Upstart's use of education and employment information in its credit underwriting model. Press Release, Consumer Fin. Prot. Bureau, CFPB Announces First No-Action Letter to Upstart Network: Company to Regularly Report Lending and Compliance Informa-

## II. Artificial Intelligence and Machine Learning

The concept of artificial intelligence emerged from the field of computer science in the 1950s and broadly refers to machine-based operations that mimic human intelligence. One of the first ideas around the concept of AI was that a computer would be understood to display intelligence when a person could not distinguish whether he or she was communicating with a human or the computer. Since then, the definition has become much broader and is used more generally to describe situations in which a computer is not given a specific set of rules to follow in order to determine an outcome.

"Machine learning," as a term, is generally viewed as a subfield of AI; we define it as the areas of mathematics, statistics, and computer science that are used to iteratively acquire knowledge and make better predictions of whatever task the computer is assigned to solve.[9]

Regardless of whether one is referring to AI or ML, the method through which a machine learns is a four-step process. First, the model builder assigns a task to the computer; second, the model builder provides the method (i.e., the algorithm) the computer should use to solve the task; third, he or she provides the measure of how the computer should evaluate its performance; fourth, the computer gets better at solving the problem through experience.

As an example, suppose that we tell the computer to attempt to determine who is likely to default on a loan, and we use credit scores as a variable to explain default. Next, we tell the computer which algorithm to use. This might be a "neural network," a "gradient boosting machine," or one of many other types of algorithms. We then tell the computer to evaluate its performance by trying to minimize the number of people who are estimated to be good risks, but who truly default ("false positives").

As a first step, the computer might assign everyone the same likelihood of default. It would then evaluate that assignment and find that it overestimates default for people with high credit scores while underestimating default for people with low credit scores. With this knowledge, it then changes its estimates of default, raising the estimates for people with low credit scores and lowering them for people with high credit scores. The computer continues the evaluation and improvement process until it reaches some preset condition or cannot make its estimates more accurate.

---

tion to the Bureau (Sept. 14, 2017), https://www.consumerfinance.gov/about-us/newsroom/cfpb-announces-first-no-action-letter-upstart-network/ [https://perma.cc/L6MN-7A8K].
9. There is a significant amount of what we view as a somewhat pedantic argument over the exact definitions of "machine learning" versus "artificial intelligence," but the definition provided above is sufficiently accepted by most practitioners.

### III. Big Data and Alternative Data

Like any predictive model, the development of ML models requires the use of data. There has been a proliferation of what has been called "big data," generally defined as information that is large in scale and complex in its interrelationships. ML methods are particularly adept at handling these large and complex datasets, being able to uncover previously unknown relationships between data elements. Importantly, however, ML does not necessarily require large amounts of data: some of these techniques can be very effective at model building with relatively small amounts of data.

"Alternative data," on the other hand, is not defined by its scale or complexity, but by its novelty. Alternative data generally refers to information used in decision-making or model building in a given context that typically is not, or historically was not, used to inform decisions in that context. A prime example is the use of alternative data in credit decisions. Until recently, information including rental payments, utility payments, occupation, social media interactions, or educational attainment were not available or rarely, if ever, used to make credit decisions. While all five of these factors may be predictive of a consumer's creditworthiness, there are reasons to think that their availability, reliability, and validity may be so correlated with protected class status that they may ultimately be unfair, raising discrimination concerns. What is important, however, is to recognize that we do not know this at the outset: the degree of unfairness may be dependent on the data, the people included in the dataset, or the other variables being considered. Ultimately, whether these data do represent unfair characteristics should be tested before they are used to make decisions that would affect consumers' lives.

Interesting examples of the value of alternative data were published in a Federal Deposit Insurance Coporation (FDIC) working paper, *On the Rise of the FinTechs—Credit Scoring using Digital Footprints*.[10] They studied whether alternative data could be predictive of default rates for a German e-commerce company.[11] Among their findings was that "the difference in default rates between customers using iOS (Apple) and Android (for example, Samsung) is equivalent to the difference in default rates between [the 50th percentile of the] credit score and the 80th percentile of the credit score."[12] They also found that customers whose email addresses contained their names were 30% less likely to default relative to others, and that customers who only used lower case letters when typing their names and addresses were more than twice as likely to default.[13]

---

10. Tobias Berg et al., *On the Rise of the FinTechs—Credit Scoring Using Digital Footprints* (Fed. Deposit Ins. Corp. Ctr. for Fin. Research, Working Paper No. 2018-04), https://www.fdic.gov/bank/analytical/cfr/2018/wp2018/cfr-wp2018-04.pdf [https://perma.cc/RAZ6-VPXX].
11. *Id*. at 2.
12. *Id*. at 3.

If the research findings regarding device ownership hold in the United States, and device ownership is correlated with protected class status, then the use of device ownership in a credit model could raise concerns about discrimination risk, particularly disparate impact. Whether it would be fair or wise for a lender to rely on the naming conventions of email addresses when determining who should get a loan is an additional open and important question. On the one hand, it does appear to be quite effective at predicting default (especially when used in conjunction with other information). However, there are many reasons why a predictive quality may not be sufficient. To provide one example, it would be easy to "game" the algorithm: if the applicant is more likely to get a loan by including their name in their email address, then it would be wise for that person to take a few minutes before applying to create a new email account. Obviously, that person's true likelihood of default has not changed as a result of changing their email address, but the algorithm will think that it has.[14]

Another question raised is that of causality, which becomes especially acute when explaining to a consumer why he or she was refused a loan. While we can surely guess why consumers who include their name in their email address might be less likely to default, the words in one's email address do not directly or even indirectly cause default. If the relationship between email address and default is real (i.e., not the result of spurious correlation), then it is more likely that both share a common cause, rather than being causally related: a person's choice of email address is at least partially driven by of the presence of some underlying factor which is causally related to poor credit outcomes. For example, since most company email addresses contain a person's name, if a person applies for a loan using a work email, then this indicates that he or she is employed. In that case, the underlying factor that is driving the predictive quality of an email address is, in fact, that the person's email address is indicative of them being currently employed, which diminishes default risk. If the proxy for employment is the only factor driving email-to-default relationship in the model then, ultimately, that means that employed customers who use personal email addresses without their names in them are being penalized more than they would be if their employment status were more directly measured.[15]

---

13. *Id.*
14. If gaming behavior were widespread and the model were refit over time, it is likely that the gameable feature would lose predictive value. So, in practice, the inclusion of the gaming variable in the model may be a short-lived phenomenon.
15. We will note that the employment-email explanation we put forward here is simply a hypothesis—we do not know if it is true. It would be possible to While name-in-email address is clearly not causally related to credit outcomes, many traditional variables used in credit models are similarly not directly causally related to credit outcomes. However, the common causal relationship underlying the traditional variables and credit outcomes tends be less tenuous than the name-in-email address example. For example, overdrawing one's checking account several times in a short time period is not likely to lead directly to default on a longer-term loan, but it is clearly indicative of the presence of financial instability that could lead to default. While this represents a relationship that is not directly causal, there is an obvious and intuitive connection to the underlying relationship between overdrawn checking accounts and a lender's justification for rejecting a customer. This is important when a lender provides a customer with an explanation for rejection (i.e., an adverse action notice), so that the customer can infer the relationship and takes steps to mitigate it. Using the hypothetical email address-employment explanation above, a consumer is unlikely to infer that their employment status was the underlying cause of rejection if the reason he or she receives is, "your name was not part of your email address."

## IV. The Positive Side of AI

Putting aside questions of whether the inclusion of particular variables would be fair or understandable, the primary value of these models and the use of reasonable alternative data is the likely improvement in predictive power relative to methods that rely on traditional data or traditional statistical methods. More predictive models (i.e., models with the ability to reliably predict good and bad outcomes) lead to more effective decision-making and better overall outcomes for both the businesses that use them and the people who are being scored with them.

Having more predictive models may lead to improved outcomes for everyone impacted by the algorithm. In credit lending, this means that a better model can lead not only to higher profits and less risk for the lenders but also help the consumers who are applying for loans. Specifically, a better ML or AI model may lead a bank to expand credit to additional customers relative to what they would have offered under a traditional credit model. This is because, as predictions get better, the risk of extreme losses decreases. As a simplified example, suppose that a bank has a risk policy such that it cannot tolerate more than a 1% chance of $10 million in losses over a year. This might mean that, using a traditional model, the bank could loan $100 million to consumers. But if an ML model can be found that has more precise predictions, then the bank may be able to lend

---

extend this list to numerous, mostly specious, possible explanations. What this further emphasizes is the need for testing and human review when using these types of alternative data.

$200 million while not changing its risk appetite (that is, a 1% chance of losing $10 million). Further, because the risk has been better quantified, it is likely that consumers will be, on average, charged lower interest rates. Confirmation of this pattern recently came through results released by the CFPB of its first No Action letter to Upstart Network, Inc. When comparing Upstart's use of alternative data and modeling to a traditional model, the CFPB reported that approvals under the alternative model were 27% higher than they would have been under traditional scoring, and that APRs were, on average, 16% lower.[16]

A further value of ML is that decisions can, in fact, be fairer as a result of removing decision-making that relies heavily on human involvement or discretion. This is because minimizing human involvement lowers the risk of human error and the influences of implicit or explicit bias. While there has rightly been concern about biased input data leading to biased predictions from AI models, one possibility that has not been extensively discussed or studied is that AI automation has the potential to ameliorate these concerns. This is because a well-built model that evaluates objective criteria will make its decisions about which variables are important based solely on the predictive quality of those variables. For instance, if a bank were to build an AI model based on historical decisions by loan officers, some of whom are biased, then the model will likely perpetuate this bias when it is first implemented. But over time, as the model reevaluates its predictions in light of true (and assumedly less biased) default outcomes, the model is likely to realize the loan officers' bias was not predictive of true behavior, and it will eventually minimize or remove the initial bias.

### V. What Does it Mean for an Algorithm to be Fair?

Despite the potential to minimize bias as discussed above, ML algorithms and the data on which they rely are not inherently bias-free. A dataset used to train an ML algorithm may, itself, reflect outcomes of a biased decision-making process (human or otherwise), and a well-intentioned model builder may create an algorithm that perpetuates the existing bias found in the data used to train the algorithm. Imprecise or poorly-built algorithms may also pick up spurious correlations in data that inadvertently lead to relatively worse outcomes for different demographic groups. Further, if protected groups are substantially underrepresented in the model's development data, and they have sufficiently different characteristics than the predominant group, then the protected class may experience biased outcomes. In consumer lending, if the disfavored demographic groups are

---

16. Patrice Ficklin & Paul Watkins, *An Update on Credit Access and the Bureau's First No-Action Letter*, Consumer Fin. Prot. Bureau (Aug. 6, 2019) https://www.consumerfinance.gov/about-us/blog/update-credit-access-and-no-action-letter/ [https://perma.cc/TB8M-Q9KD].

protected under the ECOA, the algorithm may be legally viewed as discriminatory.[17]

While the impact of human bias can be minimized, in many situations it likely cannot be completely eliminated: when building a model, a programmer must make decisions about which algorithms to deploy, supply instructions for the algorithms to follow, identify the data for training, and ultimately build a final predictive model. If there is bias in any of those steps (even unintentional), then that bias is likely to propagate through to the model's predictions.

A further complication in addressing fairness is defining what constitutes a "fair" decision. Academic work has shown that it is a far subtler issue than one would expect. For example, recent research identified twenty-one different definitions of fairness; most, if not all, of which have some intuitive appeal.[18] The research highlighted that these definitions are not internally consistent: to achieve one measure of fairness requires a trade-off where the quality of another measure of fairness deteriorates.[19] Recognizing that competing definitions of fairness cannot be concurrently satisfied, the choice of which definition to use often hinges on the context or domain in which fairness is being evaluated. This is, of course, likely to be influenced by statutory or regulatory requirements.

Recent work by Kleinberg, et al. demonstrates the likely inconsistency between conventional measures of fairness.[20] The authors consider a model that scores risk and assigns individuals to negative and positive classes (e.g., good and bad outcomes) based on observed characteristics.[21] Individuals are in one of two groups, and group membership is not a characteristic included in the predictive model.[22] Three types of fairness are considered:

---

17. The ECOA makes it unlawful to engage in three types of discrimination: overt, disparate treatment, and disparate impact. Overt discrimination is blatant discrimination on a prohibited basis (including, but not limited to, race, ethnicity, sex, or age). Disparate treatment occurs when applicants are treated differently as a result of class membership. Disparate impact occurs when a facially neutral policy or practice has a disproportionate impact on a protected class, unless the practice meets a legitimate business need that cannot reasonably be achieved as well by less discriminatory alternatives. Disparate treatment, where a lender considers or includes prohibited basis characteristics in extending or denying credit, is illegal. However, disparate impact, where there are correlations between variables in a model and the excluded prohibited basis characteristics, may be legal.
18. Arvind Narayanan, *Translation Tutorial: 21 Fairness Definitions and Their Politics*, Ass'n for Computing Mach. (Feb. 23, 2018), https://fatconference.org/2018/livestream_vh220.html.
19. *Id.*
20. Jon Kleinberg et al., Inherent Trade-Offs in the Fair Determination of Risk Scores, *in* Innovations in Theoretical Comput. Sci. Conference (2017), https://arxiv.org/pdf/1609.05807v1.pdf [https://perma.cc/H9NJ-ZMEA].
21. *Id.* at 1.
22. *Id.* at 3.

calibration within groups; balance for the negative class; and balance for the positive class.[23]

"Calibration" within groups requires that, for a given level of assessed risk, the average probabilities of classification are equal across the two groups.[24] For example, among people where the model has estimated a 20% chance of default, we see that both 20% of men and 20% of women truly do default. On the other hand, the model would not achieve calibration fairness if, among those estimated to have a 20% chance of default, we see that women only default 15% of the time, while men default 25% of the time.

"Balance for the negative class" requires that individuals who are assigned to the negative class (those who do not default) must have average risk scores that are equal across groups.[25] To be fair in terms of negative balance, average credit scores for individuals predicted not to default must be equal across both the protected and control groups. So, for people who receive a loan (because they are predicted to be less likely to default), the model would have to score both men and women with the same average (perhaps, a 5% average) chance of default.

"Balance for the positive class" is a similar concept to "balance for the negative class." But the requirement here is that there must be equivalent average scores for individuals assigned to the positive class (those who do default).[26] This means that, among the people who are rejected for a loan because they have a high expected likelihood of defaulting, both men and women must have the same average scores. For example, a model that would not show balance for the positive class is one in which Hispanics who have been rejected are predicted to have a 25% chance of default, but rejected non-Hispanic whites have a 35% chance of default.

Kleinberg et al. demonstrate that satisfying all three definitions of fairness can happen in only two cases, which are both highly unlikely in the real world.[27] The first case, defined as "perfect prediction," is when a set of observed values for characteristics used in the model are *always* associated with classification into the negative or positive class.[28] In practice, it is highly unlikely that an observed set of characteristics will always be associated with a given outcome. For instance, for a credit scoring model that relies solely on whether an individual has high or low-income, perfect prediction would require all high-income individuals to be in the same class (e.g., not default) and all low-income individuals to be in the same class (e.g., default).[29] In the real world, however, some high-income individuals do default and some low-income individuals do not. The second possible way that all three definitions can be met requires that the two groups have "equal base rates," which are defined as each group's true average probability of classification.[30] In our credit scoring example, equivalence of base rates would require that both the class and control groups have the same average default rate.[31]

---

23. *Id.*
24. *Id.* at 4.
25. A note on terminology: From the perspective of a consumer, not getting a loan or defaulting on it is, of course, a negative outcome. However, when we discuss modeling default, we are estimating the chance that the person *does* default. Thus, "does default" is referred to as the positive outcome, and "does not default" is described as the negative outcome.
26. Jon Kleinberg et al., Inherent Trade-Offs in the Fair Determination of Risk Scores, *in* Innovations in Theoretical Comput. Sci. Conference 4 (2017), https://arxiv.org/pdf/1609.05807v1.pdf [https://perma.cc/H9NJ-ZMEA].
27. *Id.* at 5.
28. *Id.*

In most applications that rely on socioeconomic data, the underlying base rates of outcomes of interest are likely to differ on the basis of subgroups—such as race, ethnicity, and sex—which means that the three fairness measures above will almost never be concurrently satisfied.[32] Further, Kleinberg et al. demonstrate that a fourth popular measure of fairness—statistical parity—is inconsistent with the calibration and balance fairness conditions when base rates differ across groups.[33] "Statistical parity" requires outcomes for the populations of subgroups to be equivalent, which is often a goal in domains where anti-discrimination laws are relevant.[34] In the context of credit-decisioning, evaluations of fairness tend to focus on statistical parity.[35]

## VI. What Can Be Done to Make AI Fairer

Despite the complexities and nuances that give rise to bias and unfairness in the use of AI models and the multiplicity of fairness goals that are being debated amongst academics and practitioners, in our work we have seen that it is possible to make AI models fairer. The focus of this approach is to take a "baseline model" that has been built without consideration of protected class status, but which shows disparate impact, and then search for alternative models that are less discriminatory than that baseline model, yet similarly predictive. While our approach is closely tied to the legal fairness standards that apply to credit, employment, and insurance provisioning, we believe that this general framework is portable to other settings, regardless of whether fairness is a legal requirement or simply a development objective. Below, we focus specifically on the framework that we use in the context of credit models.

---

29. *Id.*
30. *Id.*
31. *See id.*
32. *See id.*
33. *Id.* at 3.
34. *See id.*
35. For example, the extent to which application denial rates are equal across groups that are protected by the ECOA.

Our general approach focuses on evaluating model outcomes for disparate impact discrimination risk. The approach mirrors the legal burden-shifting test that originated in employment law and has become the relevant legal test for disparate impact discrimination.[36] Under the test: (1) plaintiff identifies an adverse impact on a protected class, (2) defendant (creditor) has the opportunity to identify a legitimate business need, and (3) plaintiff proposes a less discriminatory alternative (that reasonably achieves the same business objective). In our review, there is no distinction between the role of plaintiff and defendant. The burden-shifting test serves as a guide, and models are evaluated from the perspective of all three prongs.

The review typically starts after a model has been developed but before it has been deployed. In the context of credit and employment (and, to some extent, insurance), protected class characteristics such as race, ethnicity, and sex are explicitly excluded from the model building, during the development process, but are made available to us when conducting an assessment of disparate impact. Outcomes of the models are reviewed and if there are material differences in outcomes on the basis of protected class, we turn to consideration of legitimate business need. In banking, the model governance function is generally quite robust, so the question of whether the model has a valid business justification, (i.e., does it have the ability to predict default at some minimum level of accuracy and precision), has already been established. Consequently, the legitimate business need requirement is nearly always met. However, while the overall model may be sufficiently predictive to be considered to have a valid business justification, it is essential to review the individual variables that go into the model to ensure that they are reasonable.

Assuming that all of the variables used in the model have valid business justifications, the review then focuses on finding alternative models that are less discriminatory but similarly predictive.[37] This is where AI is likely to be better than traditional statistical modeling. Because AI models are more complex, a modeler has more ways to change or tweak the model in order to attempt to make the model less discriminatory. Further, because new methods have been developed to deal with discrimination in AI, we can address these questions head-on and find less discriminatory alternatives that preserve a baseline model's predictive power.

In traditional statistical models, the only option for finding less discriminatory models is to add or drop variables that go into the model. For instance, suppose that we find that a given credit score is the primary driver of disparate impact in the model. In our alternatives analysis we would search for other variables, perhaps including a different type of credit score, that results in the model having a lower disparate impact. In that case, before using the model to make actual credit decisions, the bank would remove the more discriminatory credit score and replace it with the less discriminatory variable we found.

---

36. *See* Griggs v. Duke Power Co., 401 U.S. 424, 439 (1971) (rejecting employer's use of intelligence tests and an education requirement as a prerequisite to employment due to the employer's failure to satisfy the burden of proving that the test demonstrated a reasonable measure of job performance.

37. If a lender were to find that a variable does not have sufficient business justification to warrant its inclusion, then it would be removed from the model.

But in AI, we are not limited just to adding or removing variables. Instead, there are many different approaches that we can try in order to find a less discriminatory alternative model. In addition to adding or dropping variables, there are methods that can keep the variable in the model but diminish the influence of its correlation with protected class status. As an example of one technique, each machine learning algorithm has what are called "hyperparameters," which are settings that tell the algorithm how to learn from the data. Changing these settings within an appropriate range can slightly change the predictions given by the algorithm. We have found that in many (but not all) cases it is possible to alter these in order to get a less discriminatory model. Hyperparameters are generally not part of traditional statistical methods, thus the option to change these to minimize bias is a benefit unique to using AI.

Other techniques can also be employed to minimize discrimination. One such method is to use what is known as "adversarial de-biasing," where two models work together to find the optimal outcome. Here, the first model predicts whatever outcome is being measured (e.g., default rates), while the second model uses those predictions to estimate the race, gender, or age of each customer.[38] The models then work together to maximize the quality of the predictions from the first model while minimizing the second model's ability to predict protected class status.[39] This has the effect of diminishing the influence of variables that are strongly correlated with protected class status.[40]

An additional technique that is similar in effect to the adversarial approach is known as "regularization." In this context, a regularized model works by balancing the predictiveness of the model and its relationship to disparate impact. As the model learns, it calculates a trade-off in the predictiveness gained relative to any increase in disparate impact. If the gain in prediction is not sufficient to overcome an increase in disparate impact, then the model will attempt to find a different path towards a predictive but less discriminatory solution.

---

38. Brian Hu Zhang et al., *Mitigating Unwanted Biases With Adversarial Learning*, *in* Artificial Intelligence, Ethics, and Society Conference (2018), http://www.aies-conference.com/wp-content/papers/main/AIES_2018_paper_162.pdf [https://perma.cc/P4AV-RCPE]. For a relatively non-technical explanation of the approach, along with an example of how well it works, *see* Hank Griffioen, *Fairness in Machine Learning with Pytorch*, GoDataDriven (May 22, 2019), https://blog.godatadriven.com/fairness-in-pytorch [https://perma.cc/V8W6-E9L3].
39. Brian Hu Zhang et al., *Mitigating Unwanted Biases With Adversarial Learning*, *in* Artificial Intelligence, Ethics, and Society (2018), http://www.aies-conference.com/wp-content/papers/main/AIES_2018_paper_162.pdf [https://perma.cc/P4AV-RCPE].
40. *See id.* (equating this as playing a zero-sum game to diminish any unduly negative effects).

As an example of how a regularized model works, suppose that the model being used is based on decision trees, where people either receive loans or are rejected as a result of various cutoffs chosen by the algorithm. In other words, if the model uses credit scores to determine who is likely to default, it will try various cutoffs of scores to see what is most predictive of default. Suppose the algorithm tries two cuts of the data: one credit score equals 600 and one credit score equals 700. When it tries the cutoff at 600, it finds that it correctly identifies 75% of defaulters and non-defaulters. However, it also finds that the adverse impact ratio (a common measure of disparate impact, where a value of 1.0 means that protected classes and control classes receive favorable outcomes at equal rates) is equal to 0.65. It then tries the cutoff at 700 and finds that it correctly identifies 73% of defaulters and non-defaulters (a decrease in accuracy of 3%). However, the adverse impact ratio has increased to 0.79. This represents a substantial improvement in disparate impact, while only minimally affecting the predictive power of the model, likely making the tradeoff worthwhile.

Finally, there is still the option of adding or removing variables, which can be made more efficient through "explainable AI," which overcomes the black-box nature of AI models, and helps to identify the importance and influence of model inputs on model predictions. These methods can be used to determine what may be causing bias in a model, and then attention can be focused on removing the biased effect of those variables. One of the chief advantages of these methodologies is that they can be automated and the results can be provided as part of the standard model review and documentation process.

## VII. Conclusion

The advent of the use of AI, ML, alternative data, and big data for decision-making holds promise but also poses risks. The scrutiny of advances in these areas has generated skepticism: will this information and these methods truly benefit individuals, or will the risk inherent in their use overshadow their value? In certain contexts, some of the benefits and risks are clear. Where there are risks, these risks can often be evaluated, managed, and mitigated.

In credit decision-making, AI, ML, and newly available data may be used to extend credit to consumers who are otherwise underserved by traditional credit decision-making methods, and to offer credit on better terms to consumers who already have access to credit. However, these new algorithmic models that rely on automated decision-making are not inher-ently bias free. While they do minimize human subjectivity, bias can arise in the model development process. Further, model developers can make decisions that introduce bias, and the data used to develop models may be inherently biased itself.

Recognizing the existence and nature of algorithmic bias, as well as the risks associated with reliance on newly available data, is a necessary step to designing and implementing approaches to ensure that these algorithmic decision-making processes are as fair as possible. Unlawful bias arising from the use of algorithms can be evaluated and mitigated, but doing so requires a deep understanding of these new methods, data, and the contexts and domains in which they are being used.

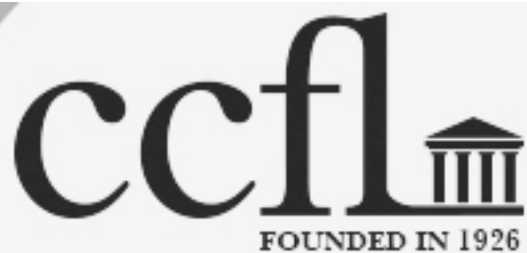

# Join the Conference on Consumer Finance Law

The Conference on Consumer Finance Law is a nonprofit corporation that encourages research in the field of consumer finance law; makes available information on the current status of laws and regulations that govern consumer finance; and, through publications and education programs, promotes the sound development of consumer finance laws that affect all types of financial institutions and businesses that offer consumer credit.

THE CONFERENCE publishes the *Consumer Finance Law Quarterly Report*, containing in-depth analysis of the most recent developments in consumer finance law.

## MEMBERSHIPS:

The *Quarterly Report* and online access to its contents are included in the annual membership dues of $95.

Sign up at ccflonline.org/membership.cfm or mail your check for $95 payable to "CCFL" to

Conference on Consumer Finance Law
P.O. Box 17981, Clearwater, FL 33762